\documentclass[structabstract]{aa}  

\newcommand{\Msun}{$M_\odot$}                               
\usepackage{graphicx}
\usepackage{txfonts}
\begin{document}
\title{Photophoretic transport of hot minerals in the solar nebula}
\authorrunning{Moudens et al.}

\author{A.~Moudens\inst{1,2}, O.~Mousis\inst{2}, J.-M.~Petit\inst{2}, G. Wurm \inst{3}, D.~Cordier \inst{1,4} and S. Charnoz \inst{5}} 
\institute{
Institut de Physique de Rennes, CNRS, UMR 6251, Universit{\'e} de Rennes 1, Campus de Beaulieu, 35042 Rennes, France
\and
Institut UTINAM, CNRS-UMR 6213, Observatoire de Besan\c con, BP 1615, 25010 Besan\c{c}on Cedex, France\\
\email{olivier.mousis@obs-besancon.fr}
\and
Faculty of Physics, University of Duisburg-Essen, Lotharstr. 1, 47048 Duisburg, Germany
\and
Ecole Nationale Sup{\'e}rieure de Chimie de Rennes, CNRS, UMR 6226, Avenue du G\'en\' eral Leclerc, CS 50837, 35708 Rennes Cedex 7, France
\and
Equipe AIM, Universit{\'e} Paris Diderot/CEA/CNRS, CEA/SAp, Centre de l'Orme Les Merisiers, 91191 Gif-Sur-Yvette Cedex, France
}

\date{Received ??; accepted ??}

  
  \abstract
{Hot temperature minerals have been detected in a large number of comets and were also identified in the samples of Comet Wild 2 that were returned by the Stardust mission. Meanwhile, observations of the distribution of hot minerals in young stellar systems suggest that these materials were produced in the inner part of the primordial nebula and have been transported outward in the formation zone of comets.}
{We investigate the possibility that photophoresis provides a viable mechanism to transport high-temperature materials from the inner solar system to the regions in which the comets were forming.}
{We use a grid of time-dependent disk models of the solar nebula to quantify the distance range at which hot minerals can be transported from the inner part of the disk toward its outer regions as a function of their size (10$^{-5}$ to 10$^{-1}$ m) and density (500 and 1000 kg\,m$^{-3}$). These models will also yield information on the disk properties (radius of the inner gap, initial mass, and lifetime of the disk). The particles considered here are in the form of aggregates that presumably were assembled from hot mineral individual grains ranging down to submicron sizes and formed by condensation within the hottest portion of the solar nebula. Our particle-transport model includes the photophoresis, radiation pressure, and gas drag.}
{Depending on the postulated disk parameters and the density of particles, 10$^{-2}$ to 10$^{-1}$ m aggregates can reach heliocentric distances up to $\sim$35 AU in the primordial nebula over very short timescales (no more than a few hundred thousand years). 10$^{-3}$ m particles follow the same trajectory as the larger ones but their maximum migration distance does not exceed $\sim$26 AU and is reached at later epochs in the disks. On the other hand, 10$^{-5}$ to 10$^{-4}$ m aggregates are continuously pushed outward during the evolution of the solar nebula. Depending on the adopted disk parameters, these particles can reach the outer edge of the nebula well before its dissipation.
}
{Our simulations suggest that irrespective of the employed solar nebula model, photophoresis is a mechanism that can explain the presence of hot temperature minerals in the formation region of comets. Comets probably had the time to trap the dust transported from the inner solar system either in their interior during accretion or in the form of shells surrounding their surface if they ended their growth before the particles reached their formation location.}


\maketitle

\section{Introduction}

Hot-temperature minerals have been detected in a large number of comets (Campins \& Ryan 1989; Crovisier et al. 2000; Sitko et al. 2004; Wooden et al. 2000, 2004, 2010) and were also identified in the samples of Comet 81P/Wild 2 that were returned by the Stardust mission (Brownlee et al. 2006). These minerals include crystalline silicates that presumably condensed in the 1200--1400 K temperature range in the solar nebula (Hanner 1999) and calcium, aluminum-rich inclusions (CAIs), which are the record of a very hot (1400--1500 K) stage of nebular evolution because they are composed of the first minerals to condense from a gas of solar composition (Grossman 1972; Jones et al. 2000). On the other hand, observations of young stellar systems show that the abundance of crystalline silicates is much higher in the inner disk than in the outer disk, but that even the outer disks show more crystalline silicates than the interstellar medium (Tielens et al. 2005). These observations then suggest that crystalline silicates, and probably also CAIs, were produced in the inner part of the primordial nebula and have been transported outward in the formation zone of comets.

A number of mechanisms has been invoked to account for the origin of these high-temperature minerals in comets. It has been proposed that shock waves in the outer solar nebula could anneal the amorphous silicates to crystallinity in situ prior to their incorporation in comets (Harker \& Desch 2002). However, the isotopic composition, minor element composition, and even the range of Fe/Si ratios measured in the dust that was returned by the Stardust spacecraft from Comet 81P/Wild 2 appear to be inconsistent with an origin by annealing of interstellar silicates in the primordial nebula (Brownlee et al. 2006). An alternative possiblity is the radial mixing induced by  turbulence which is responsible for the angular momentum transport within the primitive nebula (Shakura \& Sunyaev 1973). This turbulence favors the rapid diffusion of the different gaseous compounds and gas-coupled solids throughout the nebula. {One-dimensional (vertically averaged) diffusive transport of particles in the disk (Bockel{\'e}e-Morvan et al. 2002)} or through its surrounding layers (Ciesla 2007, 2009) has therefore been proposed to account for the presence of hot temperature minerals in the formation zone of comets. It is uncertain however whether turbulent transport suffices to explain the observations, or whether alternative physical processes are also needed. {On the other hand,} Hughes \& Armitage (2010) recently studied the outward transport of particles in the nebula via a combination of advection (inward drift of particles though interaction with gas) and turbulent diffusion in an evolving disk. These authors found that the advection of solids within the gas flow significantly reduces the outward transport efficiency for larger particles (typically a few millimeters), thereby limiting the extent of mixing uniformity that is achievable within the disk via turbulent diffusion.

An alternative transport mechanism to turbulent diffusion whose effects have been investigated in the last years in the solar nebula is photophoresis (Krauss \& Wurm 2005; Wurm \& Krauss 2006; Krauss et al. 2007; Mousis et al. 2007; Wurm et al. 2010). This effect is
based on a radiation-induced temperature gradient on the surface of a particle and the consequential nonuniform interaction with surrounding gas. When the existence of an inner gap is postulated in the disk, this latter becomes optically thin enough for particles to see the proto-Sun, but still has a reasonable gas content, which enables the photophoretic force to push dust grains outward (Mousis et al. 2007). This process provides a mechanism to transport high-temperature material from the inner solar system to the regions in which the comets were forming. Eventually, the dust driven outward in this manner will reach a region where the gas pressure and irradiation are so low that the combined outward forces of radiation pressure and photophoresis can only balance the inward drift of particles.

In this work, we use a grid of time dependent models of the solar nebula to quantify the distance range at which particles (i.e hot minerals) can be transported from the inner part of the disk toward its outer regions as a function of their size and density as well as of the disk properties (radius of the inner gap, initial mass, and lifetime of the disk). The grid of models used here allows us to consider the full range of thermodynamic conditions that might have taken place during the solar nebula's evolution. The particles considered in our model are in the form of hot mineral aggregates with sizes ranging between 10$^{-5}$ and 10$^{-1}$ m. The trajectories of particles with lower sizes are generally influenced by radiation pressure while those of particles with larger sizes begin to be mostly affected by gas drag. The aggregates are presumed to have been assembled from hot mineral individual grains ranging down to submicron sizes. We consider that these hot minerals have formed by condensation within the hottest portion of the solar nebula, well inside 1 AU (Chick \& Cassen 1997). We also show that the determination of the dust size distribution within rings observed in young circumstellar disks and their position relative to the parent star is likely to bring some constraint on the lifetime and eventually the initial mass of the disk from which they originate.

Section 2 is devoted to the description of our modeling approach, detailing the particle transport and solar nebula models employed in this work. In Section 3 we detail the disk and particle parameters employed in our different models. In Section 4 we present and analyze the trajectories of particles determined in the frame of these models. In Section 5 we show that calculations of particle trajectories induced by photophoresis can be used as a tool to determine some physical parameters of circumstellar disks. Section 6 is devoted to the discussion of the assumptions of our model.

\section{Model}
\subsection{The photophoretic force}

Any particle embedded in gas and heterogeneously heated by light feels a photophoretic force, which usually pushes it away from the light source (Krauss \& Wurm 2005; Wurm \& Krauss 2006). The force is strongly pressure-dependent and can be stronger than radiation pressure and the gravity of the Sun by orders of magnitude in the solar nebula. This mechanism induces the migration of particles ranging from micron to centimeter sized in the solar nebula under the combined action of photophoresis, radiation pressure, and gas drag, provided that the disk is sufficiently transparent (Mousis et al. 2007).

Following the approach developed by Krauss et al. (2007), we assume here that the disk's gas flow conditions are described by the Knudsen number, $Kn$, which is defined as $Kn=l/a$, where $l$ is the mean free path of the gas molecules and $a$ is the radius of the particle. If the mean free path of the gas molecules is large compared to the considered particle sizes, i.e., for $Kn>1$, then the gas flow is in the free molecular flow regime. In the contrary case, the gas flow is in the continuum regime. In these conditions, the photophoretic force $F_{ph}$ on a spherical particle, valid for both flow regimes, can be expressed as follows (Beresnev et al. 1993): 

\begin{equation}
F_{ph}=\frac{\pi}{3}a^2IJ_1 \left(\frac{\pi m_g}{2kT} \right)^{1/2} \frac{\alpha_E \psi_1}{\alpha_E+15\Lambda Kn(1-\alpha_E)/4+\alpha_E \Lambda \psi_2},
\label{Fph}
\end{equation}

\noindent where $I$ is the light flux {(power incident per area)}, $m_g$ is the average mass of the gas molecules ($3.89 \times 10^{-27} \rm kg$), $T$ is the gas temperature, and $k$ the Boltzmann constant. $J_1$ is the asymmetry factor that contains the relevant information on the distribution of heat sources over the particle's surface upon irradiation. In the following calculations, we assume $J_1=0.5$, which corresponds to the case where the incident light is completely absorbed on the illuminated side of the particle. The energy accommodation coefficient $\alpha_E$ is the fraction of incident gas molecules that accommodate to the local temperature on the particle surface and, thus, contribute to the photophoretic effect. Here, we assume complete accommodation, i.e, $\alpha_E=1$. 

The thermal relaxation properties of the particle are summarized in the heat exchange parameter $\Lambda=\lambda_{eff}/\lambda_g$, where $\lambda_g$ is the thermal conductivity of the gas and $\lambda_{eff}$ the effective thermal conductivity of the particle. For $\lambda_g$, we adopt values for molecular hydrogen for temperatures above 150K (as tabulated by Incropera \& DeWitt 2002). On the other hand, because helium has a higher thermal conductivity than hydrogen for lower temperatures, we assume that this species determines the thermal conductivity of the gas and use values taken from the compilation of Bich et al. (1990). The expression of $\Lambda$ includes the conduction of heat through the particle and the thermal emission from the particle's surface, according to

\begin{equation}
\lambda_{eff} = \lambda_p + 4\epsilon \sigma T^3a.
\label{lambdaeff}
\end{equation}

\noindent where $\lambda_p$ is the thermal conductivity of the particle supposed here to be $10^{-3}$ W\,m$^{-1}$\,K$^{-1}$ (Mousis et al. 2007), $\epsilon$ its emissivity assumed to be 1, and $\sigma$ the Stefan-Boltzmann constant. On the other hand, the functions $\psi_1$ and $\psi_2$  in Eq. (\ref{Fph}) depend only on $Kn$ in the form

\begin{equation}
\begin{array}{l}

\psi_1 = \frac{Kn}{Kn+(5\pi/18)} \left(1+\frac{2\pi^{1/2}Kn}{5Kn^2+\pi^{1/2}Kn+\pi/4} \right),
\\
\\
\psi_2 = \left(\frac{1}{2}+\frac{15}{4}Kn \right) \left(1-\frac{1.21\pi^{1/2}Kn}{100Kn^2+\pi/4} \right).

\label{psi12}
\end{array}
\end{equation}
 		
As noted by Krauss et al. (2007), an additional photophoretic force arises if the accommodation coefficients vary over the surface of the dust grain (Cheremisin et al. 2005), but we restrict the treatment to the ``classical'' photophoretic force as given in Eq. (\ref{Fph}). In the present work, we assume that Eq. (\ref{Fph}) is valid for all parts of the solar nebula and all particles.

\subsection{Balistic transport}

In a protoplanetary disk where the gas pressure (in the midplane) decreases with distance from the star, the gas is supported by a pressure gradient and rotates slower than the Keplerian velocity (Weidenschilling 1977). Solid particles are only stable on a Keplerian orbit. Therefore, interaction with the gas leads to an inward drift of solids toward the star. For particles that couple to the gas flow on timescales short compared to an orbital period, the problem {reduces to a one-dimensional (radial) calculation}. The inward drift is then induced by the fraction of gravity (residual gravity), which is not balanced by the circular motion with the sub-Keplerian gas velocity. The force, $F_{res}$, acting on a particle of mass $m_p$ due to residual gravity is given as

\begin{equation}
F_{res}=\frac{m_p}{\rho_g}\frac{dp}{dr},
\label{fres}
\end{equation}

\noindent where $\rho_g$ is the density and $p$ the pressure of the gas.
 
In addition, radiation pressure has also to be considered for at least micron-sized particles (Krauss \& Wurm 2005). The radiation pressure force can then be expressed as follow:

\begin{equation}
F_{rad}=\pi a^2 \frac{I}{c_{light}},
\label{frad}
\end{equation}
 
\noindent where $c_{light}$ is the speed of light. The sum of the outward forces (Eq. (\ref{Fph}) and Eq. (\ref{frad})) and the inward force (Eq. (\ref{fres})) gives the drift force $F_{drift}$. We treat the problem as being purely radial here because we are mostly interested in the small particles. These small particles couple to the gas on timescales much shorter than the orbital timescale, which justifies the radial treatment as outlined in Wurm \& Krauss (2006). The radial drift velocity with respect to the nebula is then estimated to be

\begin{equation}
v_{dr} = \frac{F_{ph} + F_{rad} +F_{res}}{m_p}\tau,
\label{vdrift}
\end{equation}

\noindent where $\tau$ is the gas grain coupling time and $m_p$ the mass of the considered particle.

As larger dust aggregates drift outward, they pass from a region where the continuum flow regime is valid to a region where the free molecular flow regime applies. Hence, as with the photophoretic force, we have to consider an equation describing the gas grain friction time in both regimes. It is given by

\begin{equation}
\tau=\frac{m_p}{6 \pi \eta a}C_c,
\label{tau}
\end{equation}

\noindent where $\eta$ is the dynamic viscosity of the gas. This assumes Stokes friction, which is justified because the Reynolds numbers for the drift of particles smaller than 10 cm  are well below 1. The Cunningham correction factor, $C_c$, accounts for the transition between the different flow regimes (Cunningham 1910) and is given as (Hutchins 1995)

\begin{equation}
C_c = 1+Kn \left(1.231+0.47 e^{-1.178/Kn} \right).
\label{cc}
\end{equation}

To close the set of equations, we need to determine the dynamic viscosity $\eta$ and the mean free path $l$. In the framework of the classical kinetic theory for dilute gases (see e.g. Reif 1972), these quantities are given by

\begin{equation}
\eta = \frac{1}{3} n m_g \sqrt{\frac{8 k T}{m_g \pi}} l
\label{eta}
\end{equation}

\noindent and

\begin{equation}
l = \frac{1}{\sqrt{2} n \sigma}
\label{mfp}
\end{equation}

\noindent where $n$ is the molecule number density in the gas, and $\sigma$ the collisional cross section of the gas molecules. The latter is very difficult to obtain. It is easier to find the value of the dynamic viscosity at a given temperature for H$_2$ and then use the functional form of $\eta$ in Eq.~\ref{eta} to determine its value at any temperature. We use $\eta_{0}$~=~9.0 $\times$ $10^{-6}$ Pa$\,$s at $T_0$ = 300 K (Lide 2007). Finally, inverting Eq.~\ref{eta}, one obtains the mean free path.

We note in passing that Beresnev et al. (1993) used a normalising factor of 1/2 in Eq.~\ref{eta} instead of the 1/3 that applies for three-dimensional gases, and this may slightly modify the numerical constants in Eq.~\ref{Fph}. However, the change is likely smaller than the uncertainties because of all the approximations made to solve the conservation equations in their model.

Our description of the radial transport of particles in the disk includes their drag back toward the central star by the infalling nebula flow that moves at the velocity of $v_{ac}$. In our disk model, the accretion speed ranges from a few tens of cm/s in the inner part to below one cm/s at larger distance in the early stages and substantially decreases later on. In the simplified solar nebula model presented in Section \ref{disk}, the accretion velocity is estimated to be

\begin{equation}
v_{ac} = \frac{r}{2 t_{vis}},
\label{vac}
\end{equation}

\noindent where $r$ is the distance from the Sun, $t_{vis} = \frac{1}{3 \alpha} \frac{r^2}{H^2}\frac{1}{\Omega}$ is the typical local viscous time, $H$ is the local height of the nebula, $\Omega$ is the local Keplerian frequency given by $\Omega ^2 = G M_{\sun} / r^3$ and $\alpha$ is the viscosity parameter of the disk described in Section \ref{disk}. Finally, the position of particles is integrated from the inner edge of the disk at time $t=0$ to a position $r(t_{disk})$ at the age of the disk $t_{disk}$ via

\begin{equation}
r(t_{disk}) = \int_0^{t_{disk}} (v_{dr}(r(t),t) - v_{ac}(r(t),t))$\,$dt.
\label{pos}
\end{equation}

\subsection{The protoplanetary disk}
\label{disk}

\begin{figure}
\begin{center}
\resizebox{\hsize}{!}{\includegraphics[angle=0,width=10cm]{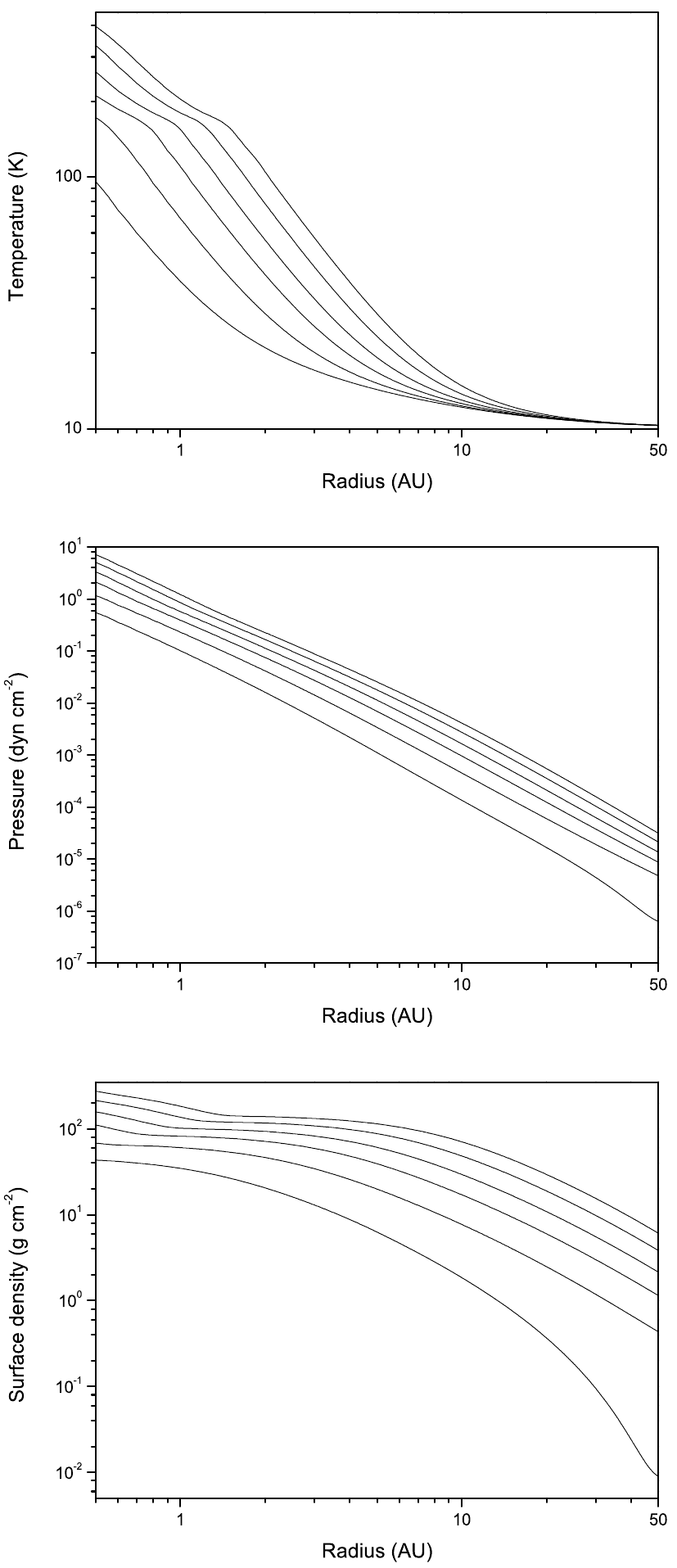}}
\caption{Temperature, pressure and surface density profiles in the midplane of the disk characterized by a mass of 0.03 \Msun~and a lifetime of 6 Myr. From top to bottom in each panel, times are 1, 2, 3, 4, 5, and 6 Myr.} 
\end{center}
\end{figure}

The structure and evolution of the protoplanetary disk is modeled as a non irradiated, 1+1D turbulent disk, following the method originally presented in Papaloizou \& Terquem (1999) and also developed by Alibert et al. (2005). The diffusion equation (see Lynden-Bell \& Pringle 1974; Papaloizou \& Lin 1995) describing the evolution of the gas surface density $\Sigma$ is consequently solved as a function of time $t$ and distance $r$ to the star:

\begin{equation}
{d \Sigma \over d t} = {3 \over r} {\partial \over \partial r } \left[ r^{1/2} {\partial \over \partial r} \nu \Sigma r^{1/2} \right] + \dot{\Sigma}_w(r)
\label{eq_diff}
,
\end{equation}

\noindent where $\Sigma$ is the surface density of mass in the gas phase in the nebula and $\nu$ the mean (vertically averaged) turbulent viscosity. Compared to the original equation, the photo-evaporation term, $\dot{\Sigma}_w(r)$, was added and is taken to be the same as in Veras \& Armitage (2004). The mean turbulent viscosity is determined from the calculation of the vertical structure of the nebula: for each radius, $r$, the vertical structure is calculated by solving the equation for hydrostatic equilibrium together with the energy equation and the diffusion equation for the radiative flux (see Papaloizou \& Terquem 1999). The local turbulent viscosity (as opposed to that averaged in the vertical direction) is computed using the standard Shakura \& Sunyaev (1973) formalism: $\nu = \alpha C_{s}^{2} /\Omega$, where $\alpha$ is a free parameter and $C_{s}$ the local speed of sound determined by the equation of state.  Using this procedure, we derived the midplane pressure and temperature as well as the mean turbulent viscosity as a function of $r$ and $\Sigma$. These laws are finally used to solve the diffusion equation (Eq. (\ref{eq_diff})) and to calculate the pressure- and temperature-dependant forces on dust grains. {Figure 1 represents the temperature, pressure and surface density profiles in the midplane of the disk characterized by a mass of 0.03 \Msun~and a lifetime of 6 Myr (see Sec. 3 for fore details) at different epochs of its evolution.}

Following the approach of Mousis et al. (2007), we consider that the disk is not optically thin and that Rayleigh scattering from molecular hydrogen is the dominant dimming effect in the nebula  (Mayer \& Duschl 2005) for temperatures below 1500~K and at wavelengths shorter than a few $\mu$m. This condition is fulfilled only after $10^5$~yr and beyond 0.5 AU in all the solar nebula models used in our calculations. For H$_2$, i.e, the dominant molecule, the Rayleigh scattering cross section is $\sigma (\lambda ) = 8.49 \times 10^{-45} / \lambda^4 \rm (cm^2)$ (Vardya 1962). Assuming the illuminating light follows a black body spectrum, the Planck mean cross section as a function of the black body temperature $T_B$ is found to be $\sigma(T_B) = 1.54 \times 10^{-42}~ T_B^4 \rm~(cm^2)$ (Dalgarno \& Williams 1962). Note that, in our case, $T_B$ is not the temperature of the nebula, but rather the effective temperature of the illuminating source, the Sun. With a disk's mean molar mass of $2.34 \rm ~g/mol$, the mass absorption coefficient is found to be $\sigma_m (T_B) = 3.96 \times 10^{-19}~T_B^4 \rm~(cm^2/g)$. {The effective temperature and the luminosity of the early Sun were taken from the ZAMS (Zero Age Main Sequence) model computed by Pietrinferni et al. (2004), which is available in the BaSTI database (http://albione.oa-teramo.inaf.it). We chose the parameters relevant for the Sun, i.e, a solar mixture of heavy elements, no overshooting, a metallicity $Z$= 0.0198, and a helium content $Y$= 0.273 ($Z$ and $Y$ together in the mass fraction). In this model, the surface temperature of the early Sun is 5652 K and its initial luminosity is 2.716 $\times$ 10$^{26}$ W. We derived $\sigma_m (5770) = 4.0 \times 10^{-4} \rm~(cm^2/g)$ from the adopted effective temperature of the early Sun.} Light becomes extinguished close to the star as a result of the high gas density, while the outer regions play only a minor role in the extinction.

\section{Choice of parameters}

We constructed a grid of nine disk models encompassing the range of thermodynamic conditions that might have taken place during the solar nebula's evolution. The three initial disk masses were fixed to 0.01, 0.03 and 0.1 \Msun~respectively, with 0.01 \Msun~corresponding to the minimum mass solar nebula (hereafter MMSN) defined by Hayashi (1981). The initial mass of each disk is integrated between 0.25 and 50 AU and the initial gas surface density is given by a power law $\Sigma \propto r^{-3/2}$, with an initial value taken to be $\Sigma (5.2 {\rm AU})$ = 100, 300, and 1000 g\,cm$^{-2}$ at 5.2 AU  for disk masses of 0.01, 0.03 and 0.1 \Msun, respectively. Here, the lifetime of the disk is governed both by viscosity and photoevaporation by the Sun or nearby stars. On the other hand, the viscosity parameter rules the accretion velocity of the disk (Eq. 12) but this latter is found to be low compared to the velocities due to photophoresis and gas drag for particles larger than 10$^{-4}$ m (see Fig. 2 for an example of particle velocities due to photophoresis, radiation pressure, residual gravity and accretion flow along their trajectories in the nebula). Here the viscosity parameter is fixed to 7 $\times$ 10$^{-3}$, i.e, a value adopted in works aiming at synthesizing different populations of planets around other stars (Mordasini et al. 2009a, 2009b) and the photoevaporation rate is varied to obtain the appropriate disk lifetimes (1, 3, and 6 Myr for each selected mass). In each case, the lifetime corresponds to the time taken for the mass of the disk (integrated until 50 AU) to decrease to 1\% of its initial value.

\begin{figure*}
\begin{center}
\resizebox{\hsize}{!}{\includegraphics[angle=0]{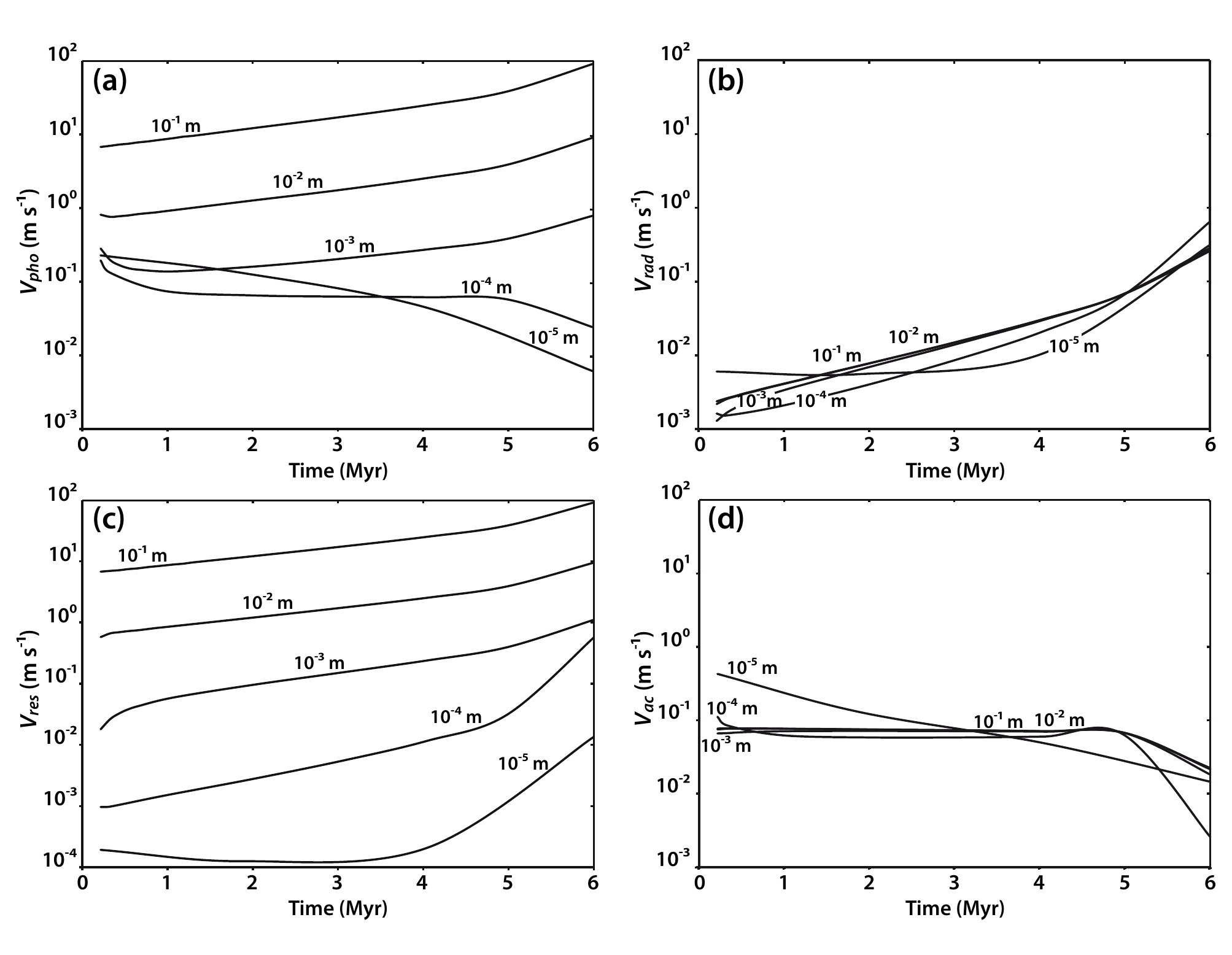}}
\caption{Velocities of particles due to photophoresis (a), radiation pressure (b), residual gravity (c), and accretion flow (d) represented as a function of time and size in the midplane of the disk characterized by a mass of 0.03 \Msun~and a lifetime of 6 Myr. The density of particles is 500 kg\,m$^{-3}$ and the radius of the inner gap is 2 AU. Position of larger particles essentially corresponds to the balance between photophoresis and residual gravity velocities. When the disk opacity is prominent, i.e., at early epochs, the position of smaller particles is mainly driven by the balance between photophoresis and accretion flow velocities. At later epochs, the position of these particles becomes ruled by the balance between all velocities.}
\end{center}
\end{figure*}

Mousis et al. (2007) have calculated the optical depth of the disk at 30 AU as a function of time. They found that even at late epochs, only $\sim$0.1\% of the Sun's radiation is available in this region. As a result, these authors found that the high extinction induced by H$_2$ Rayleigh scattering limits the outward transport of particles only to very short heliocentric distances (typically a few AU) when they are released from the innermost regions. On the other hand, particle transport can be enhanced at larger heliocentric distances when a gap is formed in the inner disk. In particular, there is a growing body of observational evidence for the existence of disks whose inner few AU are cleared or are strongly depleted of gas (D'Alessio et al. 2005; Sicilia-Aguilar et al. 2006; Espaillat et al. 2008; Pontoppidan et al. 2008; Thalmann et al. 2010). For this reason, and similar to Mousis et al. (2007), we assume here the presence of 1 and 2 AU inner gaps within the nebula during the course of its viscous evolution. Gaps are prescribed in a way independent of the structure of the disk models used in this work and their sizes remain constant with time. As shown in Sect. \ref{res}, such an inner hole is large enough to leave a reasonable fraction of the incoming light to let photophoresis work even in the outer solar system.

Particles considered in our simulations have sizes ranging between 10$^{-5}$ and 10$^{-1}$m and are assumed to be spherical and composed of olivine, with a variable porosity. Density of aggregates is varied between 500 and 1000 kg\,m$^{-3}$. The first value corresponds to the random deposition of irregular olivine particles with density of 3300 kg\,m$^{\rm-3}$, with a 15\% filling factor (Blum \& Schr{\"a}pler 2004). The second value corresponds to the average density measured in cometary interplanetary dust particles (Joswiak et al. 2007). We do not consider particles with sizes lower than 10$^{-5}$ m because their path in the nebula is essentially controlled by radiation pressure. Moreover, for objects larger than about 1 m, the radial treatment we apply does no longer hold because the gas grain friction times become comparable to the orbital period.

\section{Outward transport of hot temperature aggregates}
\label{res}

All our calculations are based on the assumption that the disk opacity is essentially caused by Rayleigh scattering and not to dust, implying that the dust size distribution in the nebula is dominated by large particles instead of small particles. In the contrary case, smallest aggregates (here 10$^{-5}$ m) would create a prominent opacity in the disk, implying that larger aggregates could only migrate outward in the wake of the small ones.

 Figures 3--6 represent the trajectories of 10$^{-5}$ to 10$^{-1}$ m aggregates in the solar nebula that were computed using the defined particle densities and a set of six disk models that are expected to encompass the range of plausible thermodynamic conditions within the solar nebula (disk masses of 1 MMSN, 3 MMSN, 10 MMSN with lifetimes of 1 or 6 Myr). At the beginning of each computation, the particles start their migration within the disk from the outer edge of the inner gap. 

Figure 3 shows that 10$^{-2}$--10$^{-1}$ m particles with densities of 500 kg\,m$^{-3}$ that migrate within a disk with a 1 AU inner gap can reach heliocentric distances ranging between {$\sim$24 and 28.1 AU}, depending on the choice of the initial mass and lifetime of the nebula. Each of these positions corresponds to an equilibrium reached at the position where the outward drift of aggregates just balances the accretion flow and in no more than a few hundred thousand years. With time, the location of these particles slightly rebounds toward the Sun until the dissipation of the disk. The figure also shows that 10$^{-3}$ m particles follow the same trajectory as the larger ones but their equilibrium position is reached at lower heliocentric distance ({$\sim$20--22.8 AU}) and at later epochs in disks owning similar input parameters. Interestingly enough, the position of smaller aggregates (10$^{-5}$--10$^{-4}$ m) continuously progresses outward during the evolution of the disks. {10$^{-5}$ m} particles can even be pushed beyond the outer edge ($\sim$50 AU) of the nebula if one selects a low-mass disk (1 MMSN) with a long lifetime (6 Myr). This is because of the strong decrease of the gas density and opacity in this model that enables the radiation pressure to push the particles at higher heliocentric distance.

Figure 4 represents the trajectories of the same particles as in Fig. 3, but for disk models with inner gaps fixed to 2 AU. Because the Rayleigh scattering through H$_2$ is strongly diminished here, all particles reach higher heliocentric distances than in the cases considered in Fig. 3, but for similar migration timescales. Thus, 10$^{-2}$--10$^{-1}$ m particles reach heliocentric distances as high as {$\sim$27--35.1 AU}, depending on the adopted parameters of the disk. In similar conditions, 10$^{-3}$ m particles are also able to reach the {$\sim$23.9--26.3 AU} distance range within the nebula. {Moreover, 10$^{-5}$ and even 10$^{-4}$ m particles reach the edge of the nebula for low mass (1 MMSN) and long lifetime (6 Myr) disk.} 

Figures 5 and 6 show the trajectories of 10$^{-5}$ to 10$^{-1}$ m aggregates with densities of 1000 kg\,m$^{-3}$ within disk models with inner gaps of 1 and 2 AU, respectively. Migration timescales remain similar to the previous cases: larger particles migrate very rapidly toward a maximum heliocentric distance while smaller ones continuously drift outward during the evolution of the disk. Because (i) the inward drift linearly depends on the mass of the aggregate (see Eq. 4) and (ii) the radial drift velocity $v_{dr}$ is inversely proportional to this quantity (see Eq. 6), all particles here migrate at lower heliocentric distances than in cases of disks based on similar parameters. Indeed, 10$^{-2}$--10$^{-1}$ m particles do not exceed {$\sim$20.2--23.6 AU ($\sim$22.7--29.7 AU)} in the case of disks with 1 AU (2 AU) inner gaps. The maximum migration distance reached by intermediary size particles (10$^{-3}$ m) becomes {$\sim$20 AU ($\sim$23.3 AU)} in the case of disks with 1 AU (2 AU) inner gaps.  {In every case, the maximum migration distance of 10$^{-4}$ m particles is several AU smaller than those of same size particles with densities of 500 kg\,m$^{-3}$. Now only 10$^{-5}$ m particles reach the outer edge of the nebula for low mass (1 MMSN) and long lifetime (6 Myr) disk, irrespective of the gap size.}\\

\begin{figure*}
\begin{center}
\includegraphics[width=15cm,angle=0]{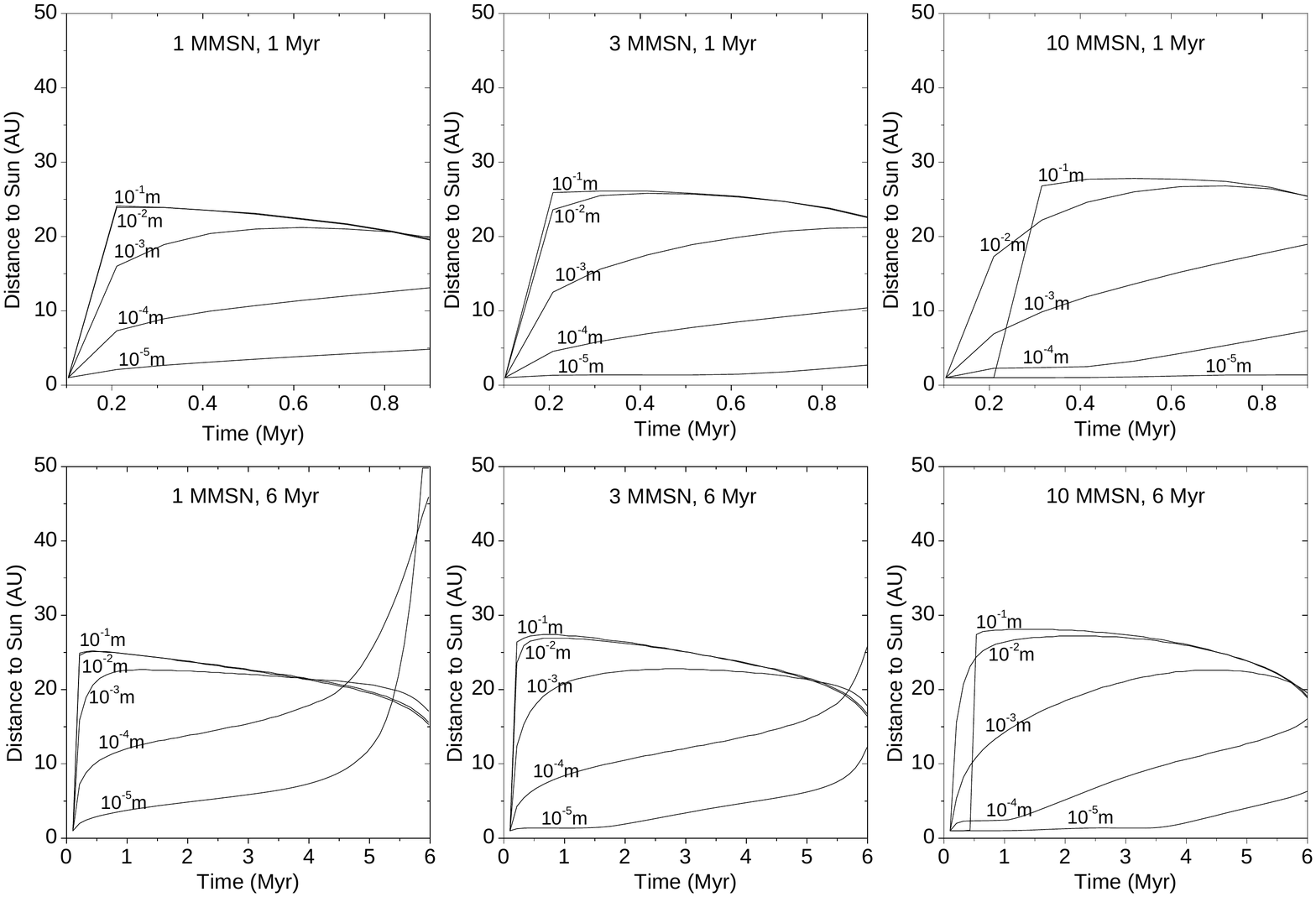}
\caption{Position of particles of size 10$^{-5}$ to 10$^{-1}$ m, as a function of time for disks with masses of 1, 3, or 10 MMSN and lifetimes of 1 or 6 Myr. The density of particles is 500 kg\,m$^{-3}$ and the radius of the inner gap is 1 AU.}
\end{center}
\end{figure*}

\begin{figure*}
\begin{center}
\includegraphics[width=15cm,angle=0]{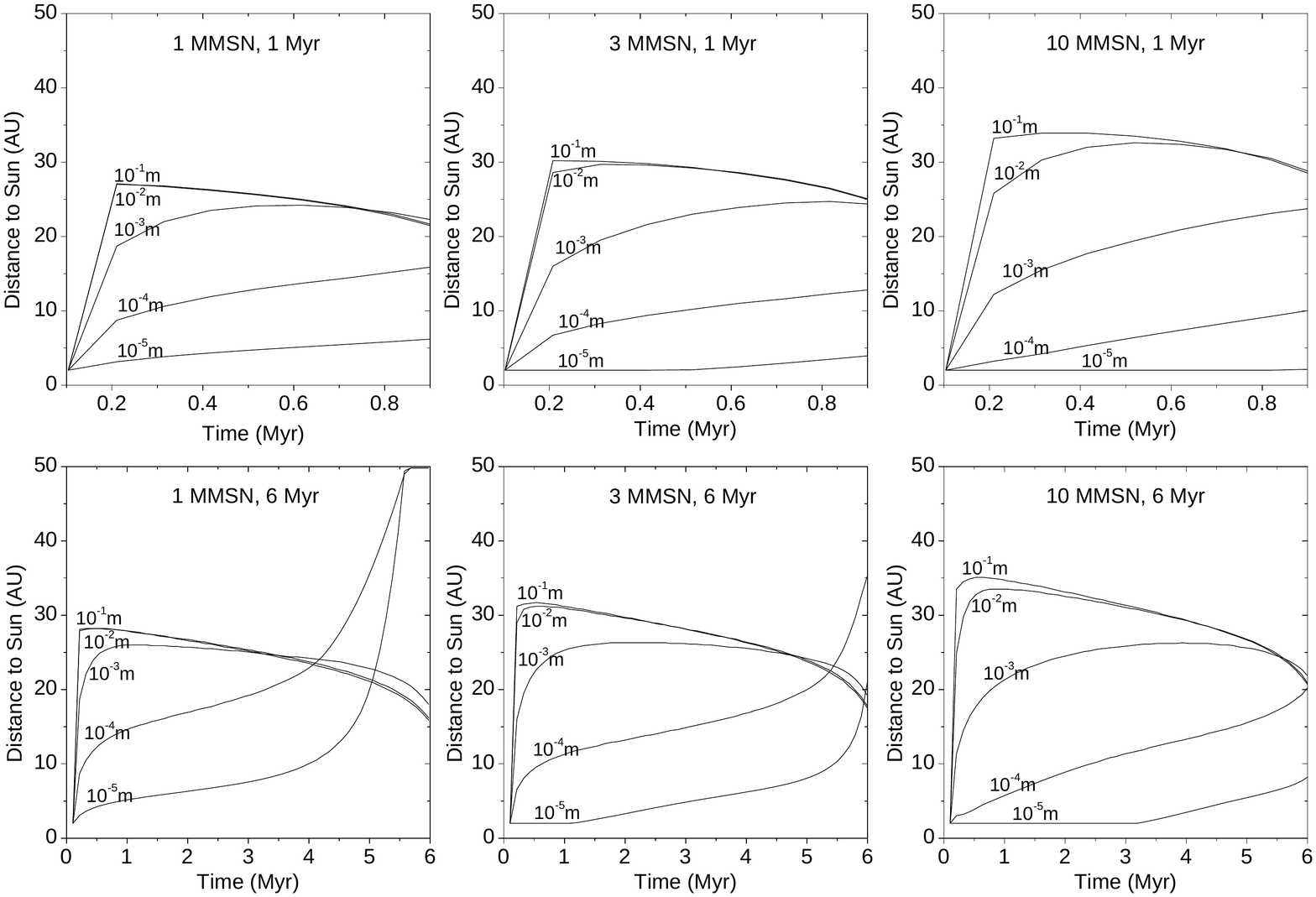}
\caption{Same as in Fig. 3, but for an inner gap radius of 2 AU.} 
\end{center}
\end{figure*}

\begin{figure*}
\begin{center}
\includegraphics[width=15cm,angle=0]{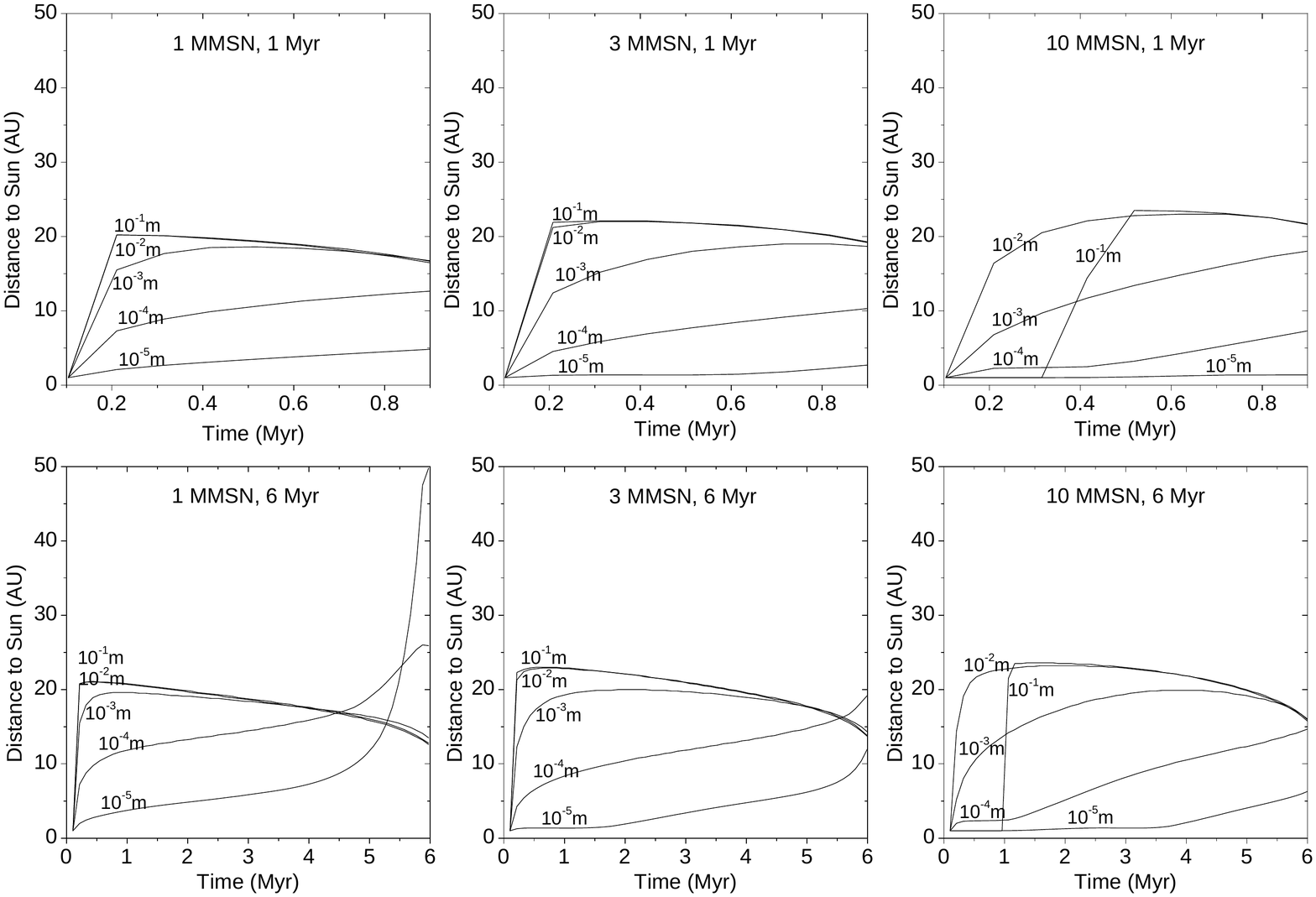}
\caption{Position of particles of size 10$^{-5}$ to 10$^{-1}$ m, as a function of time for disks with masses of 1, 3, or 10 MMSN and lifetimes of 1 or 6 Myr. The density of particles is 1000 kg\,m$^{-3}$ and the radius of the inner gap is 1 AU.} 
\end{center}
\end{figure*}

\begin{figure*}
\begin{center}
\includegraphics[width=15cm,angle=0]{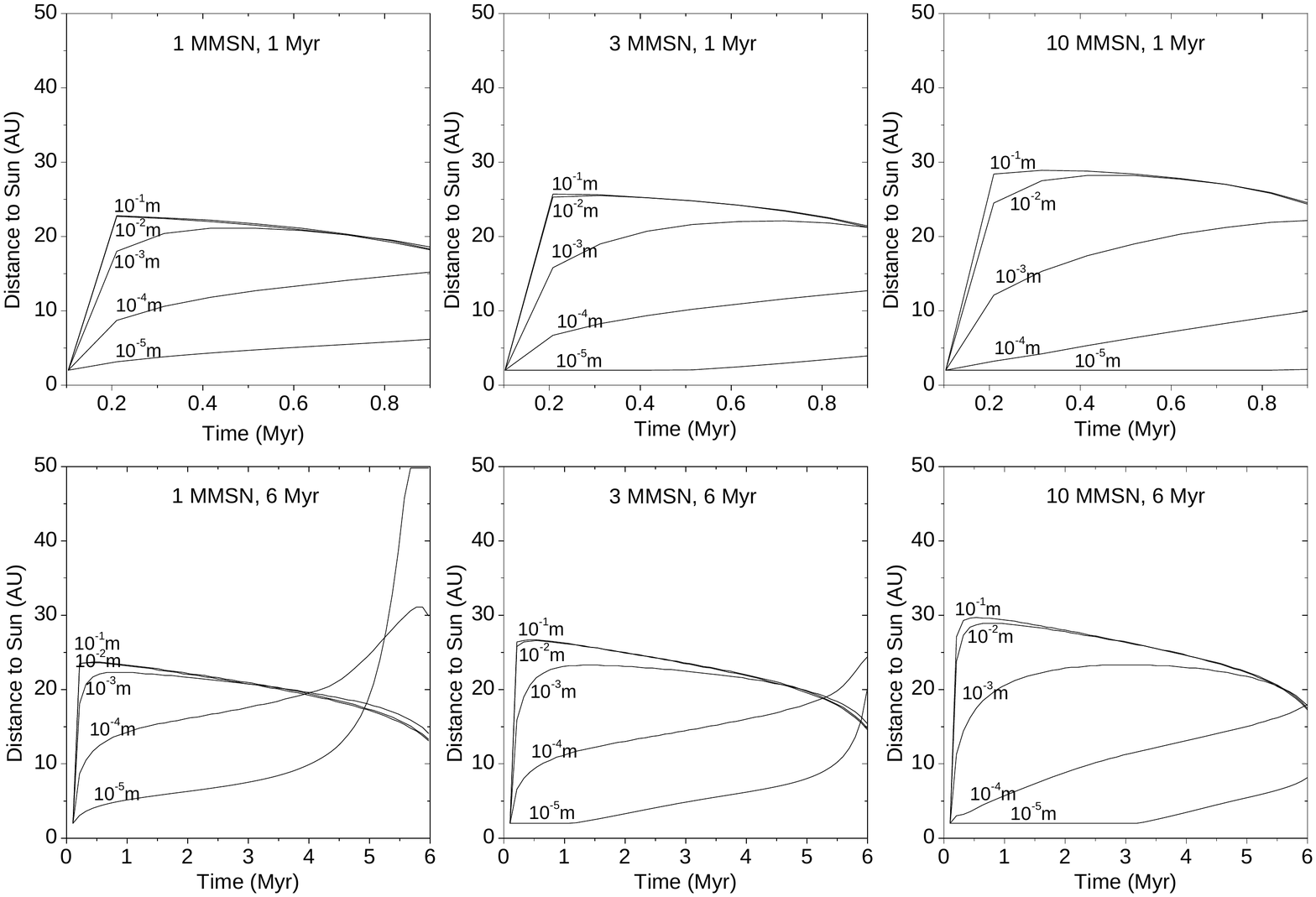}
\caption{Same as in Fig. 5 but for an inner gap radius of 2 AU.} 
\end{center}
\end{figure*}

\section{Probing the dissipation of circumstellar disks}

Particle transport through the combination of photophoresis and radiation pressure has been invoked to explain the presence of ring-shaped dust distributions in young circumstellar disks such as the one around HR 4796A (Krauss \& Wurm 2005). Here we show that in some cases, the determination of the dust size distribution within rings and their position relative to the parent star is likely to bring some constraints on the lifetime and eventually on the initial mass of the circumplanetary disk from which they originate. Indeed, Figure 7 represents the settling distances reached by particles of different sizes and with densities of 500 kg\,m$^{-3}$ at the end of the solar nebula evolution. The figure shows, for example, that ring-like structures essentially composed of 10$^{-5}$ m particles and located in the {$\sim$1.8--12 AU (3--21 AU)} distance range from the star could have formed in disks with 1 AU (2 AU) inner gap, which have short or intermediary lifetimes (here 1--3 Myr), irrespective of the initial disk's mass. In addition, same size particles located at long distance to the star, i.e., $\sim$50 AU (upper limit owing to the truncation of our model) or farther, could have formed in disks with long lifetimes (6 Myr) {and low initial masses (1 MMSN)}, irrespective of the size of the inner gap. To a lesser extent, one can also identify in Fig. 7 a relationship between the position {(in the 15--25 AU range)} of ring-like structures dominated by the settling of 10$^{-3}$ to 10$^{-1}$ m particles and the disk's lifetime and inner gap size.

\begin{figure}
\begin{center}
\resizebox{\hsize}{!}{\includegraphics[angle=0]{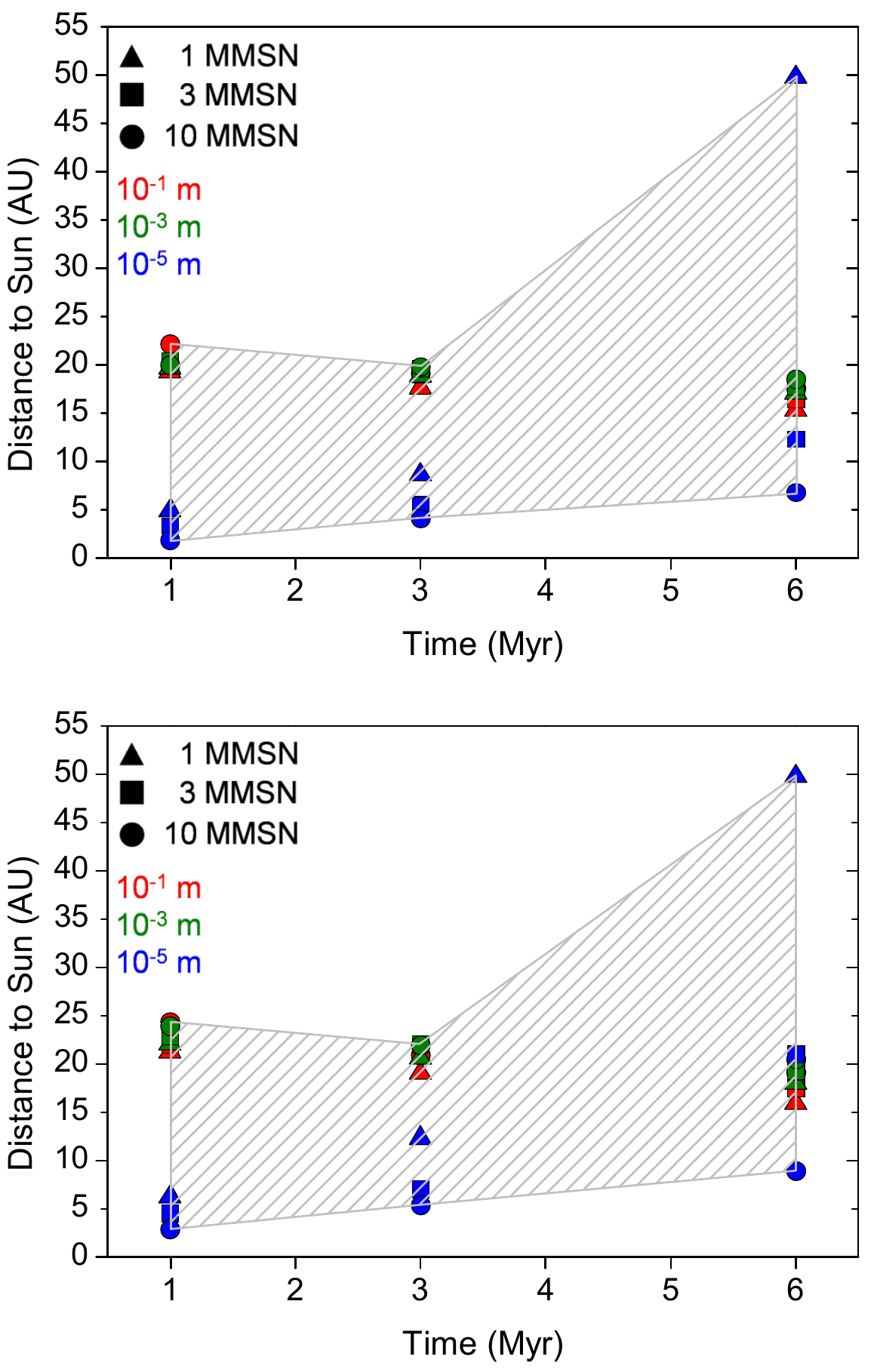}}
\caption{Heliocentric distance reached by aggregates at the end of the solar nebula evolution as a function of the disk parameters and of the particle sizes. Density of all particles is 500  kg\,m$^{-3}$ and the inner gap of the disk is 1 AU (top) and 2 AU (bottom).} 
\end{center}
\end{figure}

\section{Discussion}
\subsection{Disk's structure and evolution}
{One could argue that} the existence of an inner gap at early epochs within the nebula remains questionable. Indeed, gaps are often found in disks (i.e., transition disks) that are millions of years old. In this context, a significant offset might exist between the times of the different models used in this work and the chronology of the solar system formation and evolution that is testified by meteorite measurements or by the age of disks as inferred from luminosity studies of protostars. For these reasons, the less massive disk models used in this work that are associated to inner gaps correspond to cases that are the most consistent with the structure of transition disks. Moreover, our calculations are based on the assumption that the nebula is essentially devoid of dust, i.e., that the dust opacity is negligible. Indeed, if we assume that the smallest aggregates (10$^{-5}$ m) have created a prominent opacity in the nebula, larger aggregates could only follow the small ones and reach the formation zone of comets toward the end of the disk evolution.

Despite these caveats, the use of a set of disk models covering the whole range of plausible thermodynamic conditions that took place in the primordial nebula allows us to show that  hot-temperature minerals can drift up to heliocentric distances reaching $\sim$34 AU for the largest particles and 50 AU or beyond for the smallest ones provided that i) the existence of an inner gap is postulated within the nebula and ii) the opacity of the smallest dust particles remains negligible inside the photophoretic transport front. These simulations suggest that, irrespective of the employed solar nebula model, photophoresis is a mechanism that can explain the presence of hot-temperature minerals at early epochs of the disk's evolution in the formation region of comets (from 10 to 30 AU according to the different scenarios -- see, e.g., Horner et al. 2007 for a review). Because comets have presumably accreted within a few hundred thousand years (Weidenschilling 1997), i.e., a timescale shorter than the one probably required to form an inner gap in the disk and to allow the photophoretic transport of particles formed close to the Sun, they probably had the time to essentially trap the dust transported from the inner solar system in the form of shell surrounding their surface if their accretion ended before particles reached their formation location. Because photophoresis works heterogeneously, depending on the individual properties of a dust aggregate (composition, size, thermal and optical properties), one would expect the bulk of comets to be laden with particles of size and/or composition that would vary as a function of their accretion distance in the solar nebula. It is important to note that our calculations were made with the typical values of 0.5 and 1 for the asymmetry factor $J_1$ and the emissivity $\epsilon$. Assuming lower but still reasonable values for these parameters would not alter our conclusions. For example, if one assumes  $J_1$ = 0.4 and $\epsilon$ = 0.8 in the disk model characterized by a mass of 3 MMSN and a lifetime of 6 Myr, 10$^{-1}$--10$^{-2}$ m particles still {migrate up to $\sim$ 25.4--25.9 AU (29.5--30 AU) and 10$^{-5}$--10$^{-4}$ m particles up to $\sim$ 11--23.8 AU (18.5--33.3 AU) in the nebula owning a 1 AU (2 AU) inner gap.

\subsection{Role of turbulence}
The influence of turbulence on the particle motion has also to be considered in comparison to photophoresis. Two different cases have to be discussed. In the first case, the disk has an inner clear region and an optically thick outer region where opacity is provided by dust grains. In principle photophoresis is capable of moving the edge between optically thin and thick parts outward, thus clearing the disk inside-out from solids (see Krauss et al. 2007 for details). In this scenario, turbulent inward diffusion might counteract the outward motion of the edge. However, Krauss et al. (2007) discuss that turbulence will not prevent the outward motion of the edge. The edge will finally reach a lower heliocentric distance than in a similar situation where turbulence is negligible. The position of the edge will depend on the effective inward transport of particles by the turbulence. The second case would be more like the situation discussed in this paper. In a disk where solid particles are treated as test particles and opacity is essentially generated by the gas (Rayleigh scattering) and not by dust extinction, turbulence is not an issue. It only broadens any ring-like particle concentration because photophoresis is a directed force while turbulence is diffusive and acts statistically in both directions.

\subsection{Influence of particle rotation}

{Particle rotation is an interesting topic for photophoretic forces. Photophoresis depends on the fact that a temperature gradient is established across the particle. Usually, an illuminated particle is warmer on the bright side than on the dark side. This temperature gradient always needs a certain time to adjust to changes in the illumination. Therefore, if the particle rotates, the temperature gradient might be different from the simple case where the particle is considered at rest. Rotations might be divided into two classes. 

The first class corresponds to rotations around the direction of the incident radiation. This rotation does not change the front and back side of the particle. The temperature gradient in a reference frame fixed to the particle always stays the same. While the main component of photophoresis will still point away from the light source, the subtleties of particle morphology and composition will induce sideward components of photophoresis as well. If a particle only rotates around the direction of light, the sideward motion will oscillate but the component along the direction of light will remain constant. 

The second class corresponds to rotations around axes perpendicular to the direction of light. These rotations change front and back of the particle. For slow rotations, the temperature gradient can follow the rotation. For somewhat faster spin the gradient lags behind, meaning that the cold side will trail into the warm side and vice versa. In total this results in a certain temperature gradient directed in a direction perpendicular to the direction of light. On a given orbit, these directed forces along the orbit can accelerate or slow down the particle and make it drif inward or outward. This is a photophoretic analog to the Yarkovsky effect, which works by radiation pressure, but the photophoretic effect can be several orders of magnitudes larger. This rotation would change the calculations given here one way or the other. Wurm et al. (2010) present first experimental evidence showing this effect in microgravity experiments for rotating particles.

However, this second class of rotations can only occur temporarily and is not of general importance. To see this, we remind the reader that the idea of photophoresis is founded on the fact that particles in protoplanetary disks are embedded in a gaseous environment. This point might explicitly be noted in view of particle rotation. This simple fact is important because it implies a fundamental difference to particles in the current solar system where particles move in vacuum. While a particle in the solar system retains any arbitrary rotation state for a long time, random rotation of small particles in protoplanetary disks is rapidly damped away. For example, if a particle rotates through collisions with other particles, this rotation is gone on the order of one gas-grain friction time, which for bodies smaller than a meter is much shorter than the orbital timescale. Typical values for dust particles in the inner disk would be on the order of seconds, depending on the model and specific location in detail. Therefore, to retain a rotation, a constant torque around a given axis has to be applied to the particle. There are only two effects that can lead to torques:

\begin{enumerate}
\item Particle motion with respect to the gas, i.e., radial drift. In analogy to particles in Earth's atmosphere (like snowflakes), particle rotation will be around the drift axis. For the small particles considered here, this relative motion can be regarded as purely radial and transversal motion is not important (Weidenschilling 1977). A windmill might be an appropriate visualization for this.
\item Radiation-induced torques. In principle this might be regarded in analogy to gas drag (assuming photons instead of gas molecules) and any rotation is around the direction of illumination.
\end{enumerate}

In both cases, a torque around another axis might initially be present but this will align the particles and only the systematic torques around the drift direction or radiation direction pertain. In both cases a potential rotation axis is oriented toward the star. This will always adjust during the orbit. This rotation does not change the front and back side and does not influence radial photophoresis. Details with quantitative estimates can be found in Krauss et al. (2007).

\subsection{Prospects}

{A more realistic description of the disk's structure requires one to account for irradiation by the Sun at the disk atmosphere's surface (D'Alessio et al. 1998; Hueso \& Guillot 2005; Garaud \& Lin 2007; Cabral et al. 2010). Comparisons between irradiated and non irradiated models show that, for similar disk parameters, the midplane temperature in the outer part of the disk becomes substantially higher (up to a few dozen of K) in the first series of cases (Garaud \& Lin 2007; Cabral et al. 2010). In order to estimate the influence of irradiation on photophoretic transport, we need to know not only the temperature profile, but also the pressure and volume density in the midplane of the disk. As a first attempt, we increased the temperature by 30 K or up to 100~K for any initial temperature lower than 100 K in our nominal disk model, to mimic the difference in temperature caused by irradiation. The trajectories of larger particles (10$^{-3}$--10$^{-1}$ m) remain almost similar to those plotted in Figs. 3--6, except for the maximum distances that are 2 to 5 AU closer to the Sun. The trajectories of small particles (10$^{-5}$ m) are more affected by the temperature difference because their maximum migration distance is about half that shown in Figs.~3--6. Modifying the midplane density to keep the product $T \times \rho$ constant roughly doubles the effect.  However, both the photophoretic force and gas drag depend on the gas pressure and density and a fully consistent irradiated disk model will be needed to investigate the real influence of irradiation on the trajectories of transported particles.}

An interesting evolution of this work would also be to consider explicitly the motion of dust particles in both radial and vertical directions. Indeed, the disk's upper layers are more transparent than those close to the midplane (because of dust sedimentation), and are therefore a perfect place for photophoresis to be effective in the earliest phase of the disk's evolution and prior to the ``transition disk'' phase. At this epoch turbulence driven by active magnetorotational instability (MRI) may even be a useful ingredient because it would help a fraction of particles to be maintained above the photosphere and be transported outward thanks to both photophoeresis and radiation pressure. Indeed, computations of turbulence in MRI disks (Turner et al. 2010) have shown that turbulence is increasingly effective with scale-height, and consequently, may help to maintain particles high above. This may be a potentially interesting mechanism that would help photoporesis to be effective even in the youngest ages of the disk.}

\begin{acknowledgement}
We thank Y. Alibert for having supplied us the thermodynamic data of his disk model. O.M. and J.-M. P. acknowledge the support of CNES. G.W. acknowledges support by the DFG (SPP 1385). We acknowledge an anonymous Referee whose useful comments allowed us to strenghten our manuscript.

\end{acknowledgement}

\end{document}